\documentclass[prl,aps,twocolumn,superscriptaddress,showpacs]{revtex4}
\usepackage{bbm}
\usepackage{amssymb}
\usepackage{amsfonts}

\usepackage{graphicx,graphics,epsfig}
\usepackage{dcolumn}
\usepackage{bm}
\usepackage{amsmath}
\usepackage{verbatim}
\usepackage{color}
\usepackage{subfigure}
\usepackage{times,natbib}
\usepackage{amsmath,amsfonts,amssymb,graphics,graphicx,epsfig,color,times,natbib}

\begin{document}
\title{Einstein-Podolsky-Rosen Steerability Criterion for Two-Qubit Density Matrices}

%\author{Jing-Ling Chen \emph{et. al}}

\author{Jing-Ling Chen}
\email{chenjl@nankai.edu.cn} \affiliation{Theoretical Physics
Division, Chern Institute of Mathematics, Nankai University, Tianjin
300071, People's Republic of China}\affiliation{Centre for Quantum
Technologies, National University of Singapore, 3 Science Drive 2,
Singapore 117543}

\author{Hong-Yi Su}

\affiliation{Theoretical Physics Division, Chern Institute of
Mathematics, Nankai University, Tianjin 300071, People's Republic of
China}\affiliation{Centre for Quantum Technologies, National
University of Singapore, 3 Science Drive 2, Singapore 117543}

\author{Xiang-Jun Ye}

\affiliation{Theoretical Physics Division, Chern Institute of
Mathematics, Nankai University, Tianjin 300071, People's Republic of
China}
%\affiliation{Centre for Quantum Technologies, National
%University of Singapore, 3 Science Drive 2, Singapore 117543}

\author{Chunfeng Wu}
\affiliation{Centre for Quantum Technologies, National University of
Singapore, 3 Science Drive 2, Singapore 117543}

\author{C. H. Oh}
\email{phyohch@nus.edu.sg}
 \affiliation{Centre for Quantum
Technologies, National University of Singapore, 3 Science Drive 2,
Singapore 117543} \affiliation{Department of Physics, National
University of Singapore, 2 Science Drive 3, Singapore 117542}

\date{\today}

\begin{abstract}
We propose a sufficient criterion
${S}=\lambda_1+\lambda_2-(\lambda_1-\lambda_2)^2<0$ to detect
Einstein-Podolsky-Rosen  steering for arbitrary two-qubit density
matrix $\rho_{AB}$. Here $\lambda_1,\lambda_2$ are respectively the
minimal and the second minimal eigenvalues of $\rho^{T_B}_{AB}$,
which is the partial transpose of $\rho_{AB}$. By investigating
several typical two-qubit states such as the isotropic state,
Bell-diagonal state, maximally entangled mixed state, etc., we show
this criterion works efficiently and can make reasonable predictions
for steerability. We also present a mixed state of which
steerability always exists, and compare the result with the
violation of steering inequalities.
%Numerical results suggest that this criterion is a necessary and
%sufficient condition for demonstrating steerability of two qubits.
%The steering criterion, a sufficient condition for the
%existence of nonzero EPR steering is obtained based on partial
%transportation of 2-qubit density matrix. We apply our results to
%some models to demonstrate the utility of the criterion.
\end{abstract}


\maketitle

In 1935, Einstein-Podolsky-Rosen (EPR) questioned the completeness
of quantum mechanics based on locality and realism \cite{EPR}. Soon
after, Schr\"{o}dinger \cite{SEn} published a seminal paper defining
the notion of entanglement to describe the correlations between two
particles. Entanglement, a quantum state which cannot be separated,
is indeed the essential entity that evaluates whether a quantum
information processing can be accomplished in quantum level. The
more entanglement is, the more prowess of the resource has. Various
criteria for quantitative witnesses of entanglement
~\cite{AP1996,Horodecki,Wootters} have been proposed in recent
decades. Generally speaking, entanglement measures are mostly as
functions of density operator.

EPR steering, like entanglement, was originated from Shr{\"
o}dinger's reply to the EPR paradox to reflect the inconsistency
between quantum mechanics and local realism, and was formalized by
Wiseman, Jones, and Doherty \cite{steering1}. In the steering
scenario, for a pure entangled state held by two separated observers
Alice and Bob, Bob's qubit can be ``steered" into different
ensembles of states although Alice has no access to the qubit. Alice
tries to convince Bob that they share two systems in an entangled
state. If the systems are actually entangled, quantum mechanics
predicts that, by performing different measurements on her system,
Alice can remotely prepare different states for Bob's system. EPR
steering is commonly detected by the violation of EPR-steering
inequalities in the form of correlations
\cite{steering2,steering3,steering4,steering5,multi,Reid1,Reid2,CV1992,CV2004,entropic}.
Although many efforts have been devoted to the investigations of EPR
steering, the EPR-steering inequalities in the literatures are not
effective enough for  two-qubit systems. Therefore, it is not
possible to observe the EPR steering for some states, especially for
mixed states.
%An interesting question is whether there exists a
%criterion for the existence of nonvanishing EPR steering.
For EPR
steering, entanglement is necessary but not sufficient. By resorting
to partial transpose of density operator, entanglement can be
certified. This is understandable that the density operator contains
all the information of the state. It is hence reasonable to
anticipate a criterion based entirely on density matrix for EPR
steering witness.

In this work, we propose a criterion to detect  EPR steering of an
arbitrary  two-qubit density matrix $\rho_{AB}$. The criterion can
be obtained from the constraints on the eigenvalues of partial
transpose matrix $\rho_{AB}^{T_B}$. We list some examples to show
the utility of our criterion.

\emph{Steerability Criterion.}---Let
$\{\lambda_1,\lambda_2,\lambda_3,\lambda_4\}$ be four eigenvalues of
$\rho_{AB}^{T_B}$ in the small-to-large order [$\rho_{AB}^{T_B}$ and
$\rho_{AB}^{T_A}$ share the same
eigenvalues]. Then the  criterion for EPR steering is given by %We do the
%partial transpose on the second qubit of a two-qubit density matrix
%$\rho$, then we obtain a new non-negative matrix, of which the
%eigenvalues are generally
%$\{\lambda_1,\lambda_2,\lambda_3,\lambda_4\}$ in the small-to-large
%order. The sufficient criterion for EPR steering is
\begin{eqnarray}
\mathcal
{S}=\lambda_1+\lambda_2-(\lambda_1-\lambda_2)^2<0,\label{steering-criterion}
\end{eqnarray}
when (\ref{steering-criterion}) is satisfied, then EPR steering
exists.
%In the following, we investigate several typical entangled
%states

\emph{Example 1.}---The nonmaximal entangled state
\begin{eqnarray}
\rho_1%&=&(\cos\theta|00\rangle+\sin\theta|11\rangle)(\cos\theta\langle00|+\sin\theta\langle11|)\\
&=&\left(\begin{matrix}
\cos^2\theta&0&0&\sin\theta\cos\theta\\
0&0&0&0\\
0&0&0&0\\
\sin\theta\cos\theta&0&0&\sin^2\theta\end{matrix} \right)
\end{eqnarray}
with $\theta\in[0,\pi/4]$  always violates the CHSH inequality as
well as steering inequality given in Ref.~\cite{steering4} except
$\theta=0$. In this case, we have $\lambda\in\{-\sin\theta\cos\theta
,\sin^2\theta ,\sin\theta\cos\theta,\cos^2\theta\}$, and the
steerability criterion gives
\begin{eqnarray}
\mathcal {S}=-\frac{1}{2}\sin2\theta(1+2\sin^2\theta),
\end{eqnarray}
hence detects all the steering.

%$\lambda_1=-\sin\theta\cos\theta,\lambda_2=$.

\emph{Example 2.}---The isotropic state
\begin{eqnarray}
\rho_2&=&V\rho_0+(1-V)\frac{\textbf{1}}{4}\nonumber\\
&=&\left(\begin{matrix}
\frac{1+V}{4}&0&0&\frac{V}{2}\\
0&\frac{1-V}{4}&0&0\\
0&0&\frac{1-V}{4}&0\\
\frac{V}{2}&0&0&\frac{1+V}{4}\end{matrix} \right),
\end{eqnarray}
where $\rho_0=\frac{1}{2}(|00\rangle+|11\rangle)(\langle 00|+\langle
11|)$ is the maximal entangled state, and $\textbf{1}$ is the
four-by-four identity matrix. It has been known that the state has
the steering in the region $V\in(1/2,1]$, and no steering in
$V\in[0,1/2]$. In this case, we have
$\lambda\in\{\frac{1-3V}{4},\frac{1+V}{4},\frac{1+V}{4},\frac{1+V}{4}\}$,
and the steerability criterion gives
\begin{eqnarray}
\mathcal {S}=-\frac{1}{2}(2V-1)(1+V),
\end{eqnarray}
hence detects the critical value $V_{\rm cr}=\frac{1}{2}$.

%$\{\frac{1-3V}{4},\frac{1+V}{4},\frac{1+V}{4},\frac{1+V}{4}\}$
%
%$\lambda_1=\frac{1-3V}{4},\lambda_2=\frac{1+V}{4}$.

\emph{Example 3.}---%The combination of the maximal entangle states
The Bell-diagonal state
\begin{eqnarray}
\rho_3&=&V|\Psi^{+}\rangle\langle\Psi^{+}|+(1-V)|\chi^{+}\rangle\langle\chi^{+}|\nonumber\\
&=&\left(\begin{matrix}
\frac{V}{2}&0&0&\frac{V}{2}\\
0&\frac{1-V}{2}&\frac{1-V}{2}&0\\
0&\frac{1-V}{2}&\frac{1-V}{2}&0\\
\frac{V}{2}&0&0&\frac{V}{2}\end{matrix} \right)
\end{eqnarray}
violates the steering inequality in Ref.~\cite{steering4} except
$V=\frac{1}{2}$. In this case, we have
$\lambda\in\{V-\frac{1}{2},\frac{1}{2}-V,\frac{1}{2},\frac{1}{2}\}$,
and the steerability criterion gives
\begin{eqnarray}
\mathcal {S}=-(1-2V)^2,
\end{eqnarray}
which recovers the same result.

%$\{V-\frac{1}{2},\frac{1}{2}-V,\frac{1}{2},\frac{1}{2}\}$
%
%$\lambda_1=V-\frac{1}{2},\lambda_2=\frac{1}{2}-V$.

\emph{Example 4.}---The nonmaximal entangle state with color noise
\begin{eqnarray}
\rho_4%&=&(\cos\theta|00\rangle+\sin\theta|11\rangle)(\cos\theta\langle00|+\sin\theta\langle11|)\\
&=&\left(\begin{matrix}
V\cos^2\theta+\frac{1-V}{2}&0&0&V\sin\theta\cos\theta\\
0&0&0&0\\
0&0&0&0\\
V\sin\theta\cos\theta&0&0&V\sin^2\theta+\frac{1-V}{2}\end{matrix}
\right)
\end{eqnarray}
with $\theta\in[0,\pi/4]$ always violates CHSH inequality as well as
the steering inequality in Ref.~\cite{steering4} except
$V=\theta=0$. In this case, we have
$\lambda\in\{-V\sin\theta\cos\theta,V\sin\theta\cos\theta,\frac{1}{2}(1-V\cos2\theta),\frac{1}{2}(1+V\cos2\theta)\}$,
and the steerability criterion gives
\begin{eqnarray}
\mathcal {S}=-V^2\sin^22\theta,
\end{eqnarray}
which recovers the same result.

%$\{-V\sin\theta\cos\theta,V\sin\theta\cos\theta,\frac{1}{2}(1-V\cos2\theta),\frac{1}{2}(1+V\cos2\theta)\}$
%
%$\lambda_1=-V\sin\theta\cos\theta,\lambda_2=$.

\emph{Example 5.}---The maximally entangled mixed state (MEMS)
\begin{eqnarray}
\rho_5=\left(\begin{matrix}
g(\gamma)&0&0&\gamma/2\\
0&1-2g(\gamma)&0&0\\
0&0&0&0\\
 \gamma/2&0&0&g(\gamma)\end{matrix} \right),
\end{eqnarray}
with $g(\gamma)=1/3$ for $\gamma\in[0,2/3]$ and $g(\gamma)=\gamma/2$
for $\gamma\in[2/3,1]$. It violates the 10-setting steering
inequality in Ref.~\cite{steering4} for $\gamma\geq 0.6029$. In the
case of $\gamma\in[0,2/3]$, we have
$\lambda\in\{\frac{1-\sqrt{1+9\gamma^2}}{6},\frac{1}{3},\frac{1}{3},\frac{1+\sqrt{1+9\gamma^2}}{6}\}$,
and the steerability criterion gives
\begin{eqnarray}
\mathcal {S}=\frac{1}{36}(16-9\gamma^2-8\sqrt{1+9\gamma^2}),
\end{eqnarray}
which predicts the critical value
\begin{eqnarray}
 \gamma_{\rm
cr}=\frac{2}{3}\sqrt{2(6-\sqrt{33})}\simeq0.4765.
\end{eqnarray}

%$\{\frac{1-\sqrt{1+9\gamma^2}}{6},\frac{1}{3},\frac{1}{3},\frac{1+\sqrt{1+9\gamma^2}}{6}\}$
%
%$\lambda_1=\frac{1-\sqrt{1+9\gamma^2}}{6},\lambda_2=\frac{1}{3}$.
%
%$\{\frac{1}{2} (1-\gamma -\sqrt{1-2 \gamma +2 \gamma
%^2}),\frac{\gamma}{2},\frac{\gamma}{2},\frac{1}{2} (1-\gamma
%+\sqrt{1-2 \gamma +2 \gamma ^2})\}$
%
%$\lambda_1=\frac{1}{2} (1-\gamma -\sqrt{1-2 \gamma +2 \gamma
%^2}),\lambda_2=\frac{\gamma}{2}$

\emph{Example 6.}---The  state
\begin{eqnarray}
\rho_6=\left(\begin{matrix}
g&0&0&\gamma/2\\
0&1/2-g&0&0\\
0&0&0&0\\
 \gamma/2&0&0&1/2\end{matrix} \right),
\end{eqnarray}
with $g=4/9$ for $\gamma\in[0,2\sqrt{2}/3]$. It violates the
10-setting steering inequality in Ref.~\cite{steering4} for
$\gamma\geq 0.2564$. In this case, we have
$\lambda\in\{\frac{1}{36}(1-\sqrt{1+324\gamma^2}),\frac{1}{36}(1+\sqrt{1+324\gamma^2}),4/9,1/2\}$,
and the steerability criterion gives
\begin{eqnarray}
\mathcal {S}=\frac{17}{324}-\gamma^2,
\end{eqnarray}
which predicts the critical value
\begin{eqnarray}
 \gamma_{\rm
cr}=\frac{\sqrt{17}}{18}\simeq0.2291.
\end{eqnarray}
\begin{figure}[tbp]
\includegraphics[width=80mm]{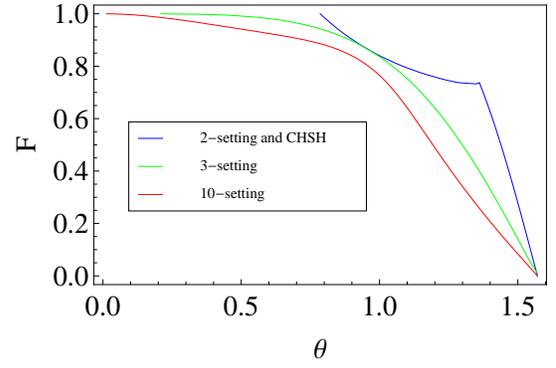}\\
\caption{(Color online) Quantum predictions of steering
inequalities. The region above the blue line is steerable detected
by the two-setting steering inequality as well as CHSH inequality.
The region above the green and red lines are respectively  steerable
detected by the three- and ten-setting steering inequalities.}
\label{fig1}
\end{figure}

\begin{figure}[tbp]
\includegraphics[width=75mm]{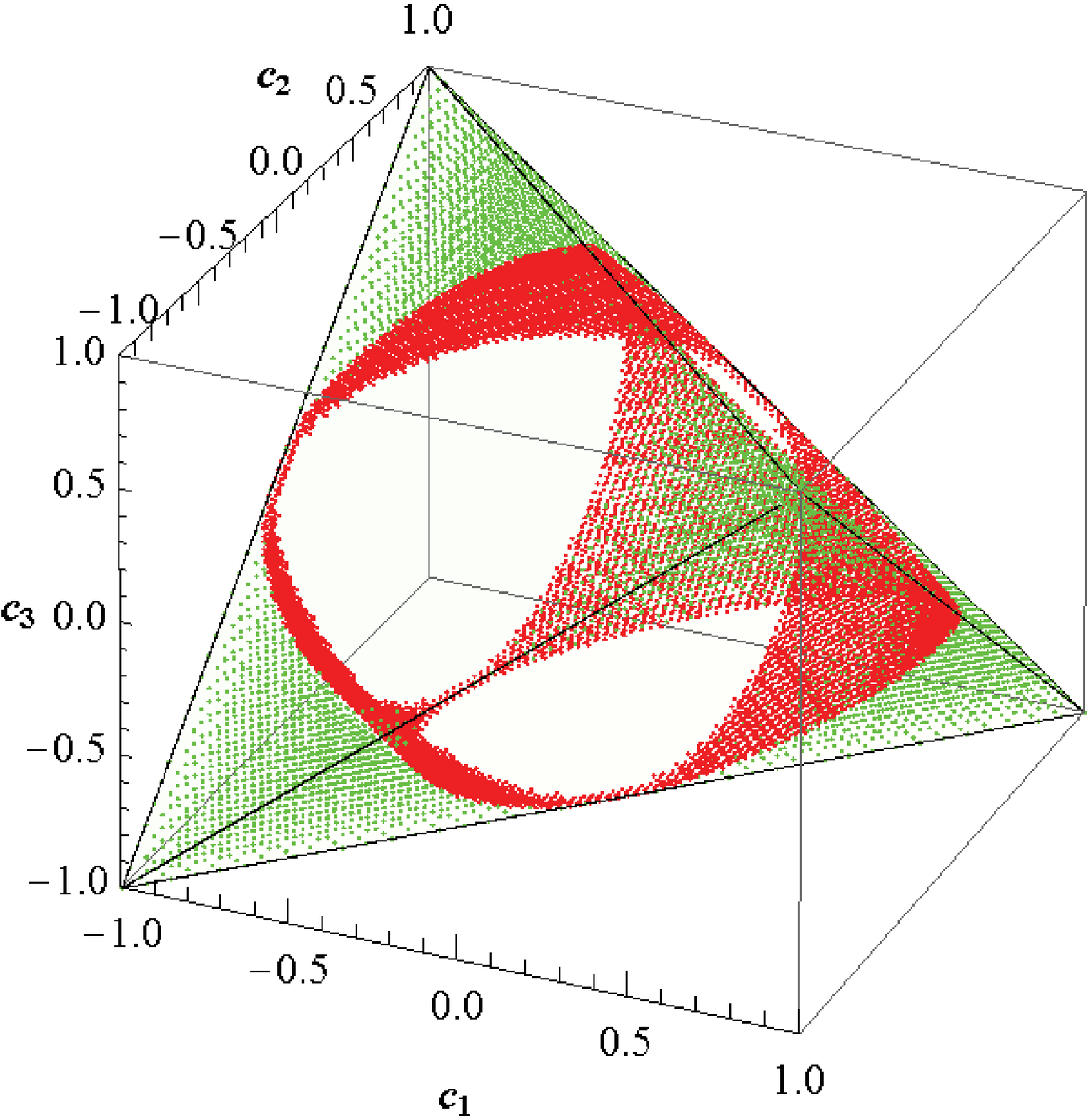}\\
\caption{(Color online) Comparison of detective ability of
steerability criterion and the steering inequality for $r=s=0$. The
tetrahedron is the set of all two-qubit states. Four vertices
represent four Bell states, respectively. The states with $\mathcal
{S}=0$ locate on the red surface, inside which are states with
$\mathcal {S}>0$, outside which are states with $\mathcal {S}<0$.
The green points represent the states that violate the ten-setting
steering inequality. Numerical result shows that the green points
are always located in the volume between the red surface and the
polytope (tetrahedrron) defined by the four vertices (Bell states).
}\label{fig2}
\end{figure}
\begin{figure}[tbp]
\includegraphics[width=75mm]{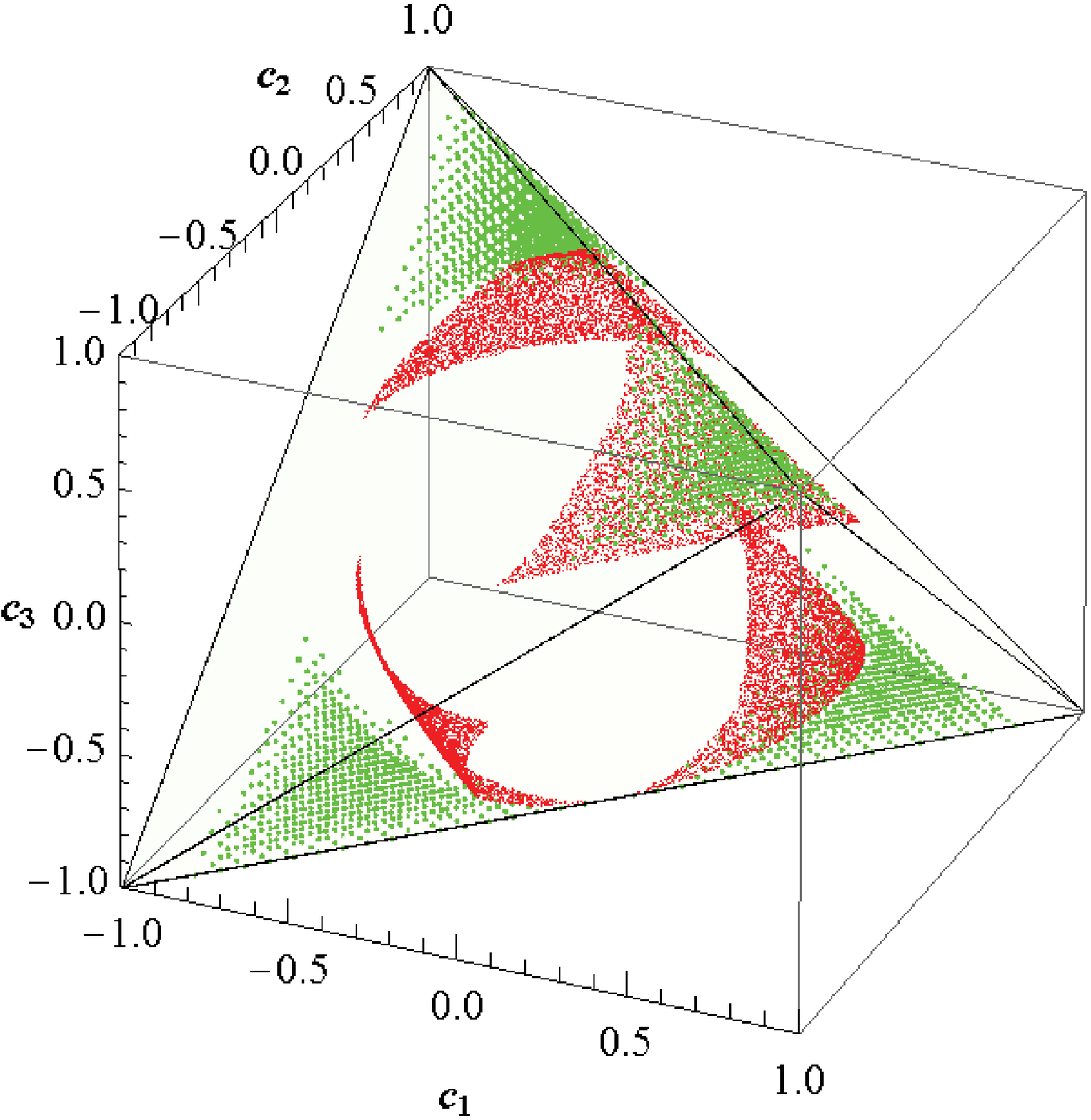}\\
\caption{(Color online) Comparison of detective ability of
steerability criterion and the steering inequality for
$|r|=|s|=1/2$.
The tetrahedron is the set of all two-qubit states.
Four vertices represent four Bell states, respectively. The states
with $\mathcal {S}=0$ locate on the red surface, inside which are
states with $\mathcal {S}>0$, outside which are states with
$\mathcal {S}<0$. The green points represent the states that violate
the ten-setting steering inequality. Numerical result shows that the
green points are always located in the volume between the red
surface and the polytope (tetrahedrron) defined by the four vertices
(Bell states).
}\label{fig3}
\end{figure}

\emph{Example 7.}---The  state
\begin{eqnarray}
\rho_7=\left(\begin{matrix}
\frac{1-\cos\theta}{2}F&0&0&\frac{\sin\theta}{2}F\\
0&\cos\theta F&0&0\\
0&0&0&0\\
\frac{\sin\theta}{2}F&0&0&1-\frac{1+\cos\theta}{2}F\end{matrix}
\right),
\end{eqnarray}
with $F\in[0,1]$ and $\theta\in[0,\pi/2]$. The concurrence of the
state is given by
\begin{eqnarray}
\mathcal {C}=\sqrt{2F(1-|F(1+\cos\theta)-1|)\sin^2\frac{\theta}{2}},
\end{eqnarray}
which vanishes only when $F=0$ or $\theta=0$. In this case,
$\lambda\in\{-F\sin^2\frac{\theta}{2},F\sin^2\frac{\theta}{2},F\cos^2\frac{\theta}{2},1-F\cos^2\frac{\theta}{2}\}$,
we have $\lambda_1=-\lambda_2=-F\sin^2\frac{\theta}{2}$, and the
steerability criterion gives
\begin{eqnarray}
\mathcal {S}=-4F^2\sin^4\frac{\theta}{2},
\end{eqnarray}
which predicts that steering always exists. We compare the above
result with the violation of the steering inequalities and CHSH
inequality (see Fig.~\ref{fig1}).

Any two-qubit  state can be written in the following form
\begin{equation}
   \rho_{AB}^{\phantom{2}} = \frac{1}{4} ( {\bf 1} \otimes {\bf 1} +
   {\vec \sigma}^A \cdot {\bf u} \otimes {\bf 1}+
   {\bf 1}\otimes {\vec \sigma}^B \cdot {\bf v}+
   \sum_{i,j=1}^3 \beta_{ij} \sigma_i^A \otimes \sigma_j^B ),
 \end{equation}
where ${\bf u}$ and ${\bf v}$ are Bloch vectors for particles A and
B, respectively; $\beta_{ij}$ are some real numbers. Particularly,
we take ${\bf u}=(0,0,r)$, ${\bf v}=(0,0,s)$ and
$\beta_{ij}=c_{i}\delta_{ij}$, then we obtain the five-parameter
$X$-state as
\begin{equation}
   \rho_{AB}^{\phantom{2}} = \frac{1}{4} ( {\bf 1} \otimes {\bf 1} +
   { r}\;{\sigma_3}^A   \otimes {\bf 1}+
   {\bf 1}\otimes { s}\;{ \sigma_3}^B +
   \sum_{i=1}^3 c_i \sigma_i^A \otimes \sigma_i^B ).
 \end{equation}
In Fig.~\ref{fig2} and Fig.~\ref{fig3} we compare the detective
ability of steerability criterion and the ten-setting steering
inequality.

%\begin{eqnarray}
%\rho=\frac{1}{4}(\textbf{1}\otimes\textbf{1}+r\sigma\otimes
%\textbf{1}+),
%\end{eqnarray}

%It violates the steering inequalities in Ref.~\cite{steering4} in
%the region shown in Fig..,%\;\lambda_2=1-f\cos^2\frac{\theta}{2}

%\vspace{30mm}
%
%\emph{EPR Steering inequalities.}---In the following are EPR
%steering inequalities with different settings.
%
%\emph{2-setting}:
%\begin{eqnarray}
%S_2=\frac{1}{2}\sum_{i=1}^2\langle A_iB_i\rangle\leq C_2,
%\end{eqnarray}
%\emph{3-setting}:
%\begin{eqnarray}
%S_3=\frac{1}{3}\sum_{i=1}^3\langle A_iB_i\rangle\leq C_3,
%\end{eqnarray}
%\emph{4-setting}:
%\begin{eqnarray}
%S_4=\frac{1}{4}\sum_{i=1}^4\langle A_iB_i\rangle\leq C_4,
%\end{eqnarray}
%\emph{6-setting}:
%\begin{eqnarray}
%S_6=\frac{1}{6}\sum_{i=1}^6\langle A_iB_i\rangle\leq C_6,
%\end{eqnarray}
%\emph{10-setting}:
%\begin{eqnarray}
%S_{10}=\frac{1}{10}\sum_{i=1}^{10}\langle A_iB_i\rangle\leq C_{10},
%\end{eqnarray}
%with upperbound
%$C_2=\frac{1}{\sqrt{2}}\simeq0.7071,\;C_3=\frac{1}{\sqrt{3}}\simeq0.5773,\;C_4=\frac{1}{\sqrt{3}}\simeq0.5773,\;C_6\simeq
%0.5393,\;C_{10}\simeq0.5236$.
%

In summary, the criterion is proposed due to numerical observation,
which works efficiently for detecting EPR steering of two-qubit
density matrix. Similar to PPT criterion for detecting entanglement,
the steerability criterion may also work as a necessary condition
for demonstrating steerability of two qubits. It would be
significant to derive the steerability criterion from analytic
approach, such as positive maps, and then place  it on a firmer
foundation.

% the
%numerical result suggests it is a necessary and sufficient condition
%to detect two-qubit EPR steering.

%The criterion is simple and experimentally testable, since it can be
%evaluated conveniently from quantum state tomography. Thus, we
%provide a direct criterion to verify the presence of EPR-steering
%for an arbitrary $2$-qubit quantum state. The criterion provides a
%novel method for quantifying EPR steering, and it offers another way
%to test the existence of EPR steering experimentally. Especially for
%mixed states, the known EPR-steering inequalities not strong enough,
%while with the criterion, the EPR steering can be revealed.

J.L.C. is supported by National Basic Research Program (973 Program)
of China under Grant No. 2012CB921900 and NSF of China (Grant Nos.
10975075 and 11175089). This work is also partly supported by
National Research Foundation and Ministry of Education, Singapore
(Grant No. WBS: R-710-000-008-271).


\begin{thebibliography}{99}


\bibitem{EPR} A. Einstein, B. Podolsky, and N. Rosen, Phys. Rev. \textbf{47}, 777 (1935).

\bibitem{SEn} E. Schr{\"o}inger and M. Born, Math. Proc. Cambridge Philos. Soc. \textbf{31}, 555 (1935);
E. Schr{\"o}inger and P. A. M. Dirac, Math. Proc. Cambridge Philos.
Soc. \textbf{32}, 446 (1936).

\bibitem{AP1996} A. Peres, Phys. Rev. Lett. \textbf{77}, 1413 (1996).

\bibitem{Horodecki} M. Horodeckia, P. Horodeckib, and R. Horodeckic,
Phys. Lett. A \textbf{223}, 1 (1996); P. Horodecki, Phys. Lett. A \textbf{232}, 333 (1997);
M. Horodecki, P. Horodecki, and R. Horodecki, Phys. Rev. Lett \textbf{80}, 5239 (1998).

\bibitem{Wootters} W. K. Wotters, Phys. Rev. Lett. \textbf{80}, 2245 (1997).

\bibitem{steering1} H. M. Wiseman, S. J. Jones, and A. C. Doherty, Phys. Rev. Lett. \textbf{98}, 140402
(2007).

\bibitem{steering2} S. J. Jones, H. M. Wiseman, and A. C. Doherty, Phys. Rev. A \textbf{76}, 052116 (2007).

\bibitem{steering3} E. G. Cavalcanti, S. J. Jones, H. M. Wiseman, and M. D. Reid, Phys. Rev. A \textbf{80}, 032112 (2009).

\bibitem{steering4} D. J. Saunders, S. J. Jones, H. M.Wiseman and G. J. Pryde, Nature Phys. \textbf{6},
845 (2009).

\bibitem{steering5} Q. Y. He, P. D. Drummond, and M. D. Reid, Phys. Rev. A \textbf{83}, 032120 (2011).

\bibitem{multi} E. G. Cavalcanti, Q. Y. He, M. D. Reid, H. M. Wiseman
, Phys. Rev. A \textbf{84}, 032115 (2011).


%\bibitem{Higgs} P. W. Higgs, J. Phys. A \textbf{12}, 309--323 (1979).

\bibitem{Reid1} M. D. Reid, Phys. Rev. A \textbf{40}, 913 (1989).

\bibitem{Reid2} M. D. Reid, P. D. Drummond, W. P. Bowen, E. G. Cavalcanti, P.
K. Lam, H. A. Bachor, U. L. Anderson, and G. Leuchs, Rev. Mod. Phys.
\textbf{81}, 1727 (2009).

\bibitem{CV1992} Z. Y. Ou, S. F. Pereira, H. J. Kimble, and K. C. Peng, Phys.
Rev. Lett. \textbf{68}, 3663 (1992).

\bibitem{CV2004} J. C. Howell, R. S. Bennink, S. J. Bentley, and R. W. Boyd,
Phys. Rev. Lett. \textbf{92}, 210403 (2004).

\bibitem{entropic} S. P. Walborn, A. Salles, R. M. Gomes, F. Toscano, and P. H. Souto
Ribeiro, Phys. Rev. Lett. \textbf{106}, 130402 (2011).








\end{thebibliography}
\end{document}